# Transferable Water Potentials Using Equivariant Neural Networks

*Tristan Maxson, Tibor Szilvási\**

*\*Corresponding author. Email: tibor.szilvasi@ua.edu*

**Abstract:**

Machine learning interatomic potentials (MLIPs) are an emerging modeling technique that promises to provide electronic structure theory accuracy for a fraction of its cost, however, the transferability of MLIPs is a largely unknown factor. Recently, it has been proposed (*J. Chem. Phys.*, **2023**, *158*, 084111) that MLIPs trained on solely liquid water data cannot describe vapor-liquid equilibrium while recovering the many-body decomposition analysis of gas-phase water clusters, as MLIPs do not directly learn the physically correct interactions of water molecules, limiting transferability. In this work, we show that MLIPs based on an equivariant neural network architecture trained on only 3,200 bulk liquid water structures reproduces liquid-phase water properties (e.g., density within 0.003 g/cm$^3$ between 230 and 365 K), vapor-liquid equilibrium properties up to 550 K, the many-body decomposition analysis of gas-phase water cluster up to six-body interactions, and the relative energy and the vibrational density of states of ice phases. This study highlights that state-of-the-art MLIPs have the potential to develop transferable models for arbitrary phases of water that remain stable in nanosecond-long simulations.

TOC Graphic:

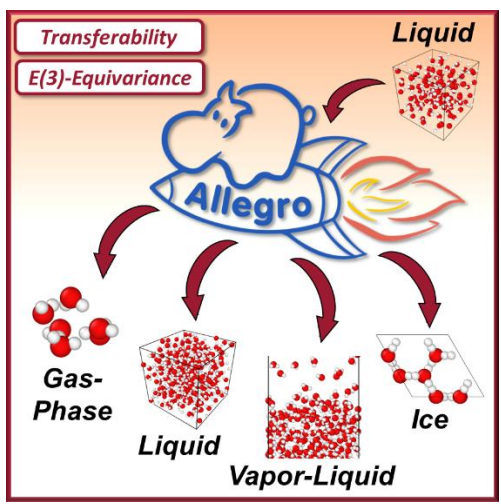

Water plays a pivotal role across various chemical, biological, and environmental processes; however, understanding the effect of water on such processes requires accurate models and long timescales for proper equilibration of water[1-6]. The computational costs render first principles/*ab initio* modeling by density functional theory (DFT)/coupled cluster theory (CC) impractical for large systems. An affordable alternative is to fit classical force fields (FFs, e.g., TIP3P[7], SPC/E[8], etc.) to DFT data that can perform relatively well for the fitted properties of water, however, transferability to other properties and phases is a general issue[9-11]. A state-of-the-art water potential, MB-Pol,[12-16] has been developed using CC data of gas-phase water clusters to provide transferable and accurate simulations of water[17]. MB-Pol includes additional computational complexity (and recently machine learned terms) compared to classical FF terms, reducing speed greatly compared to classical FFs, but recovers CC-quality accuracy[18] and general transferability between water phases[19, 20] and properties[21]. Machine learning interatomic potentials[22-26] (MLIPs) have recently emerged as a flexible new method to fit existing data and reproduce the properties of training data with high accuracy[27-29]. MLIPs promise a flexible and affordable path to developing precise and transferable water models, at a computational cost that is more affordable than DFT, CC, or complex potentials such as MB-Pol.

The development of a transferable MLIP using only liquid-phase MB-Pol water data was previously attempted by Zhai et al.[30] and more recently Muniz et al.[31] both using DeePMD[32]. While these studies showed qualitatively correct trends in liquid-phase properties like density, they failed to quantitatively match MB-Pol results over the entire temperature range studied. Difficulties in training accurate and transferable MLIPs are rationalized as MLIPs were unable to learn the fundamental intra- and intermolecular interactions of water solely from liquid-phase data. To directly probe intermolecular interactions, Zhai et al.[30] performed many-body decomposition (MBD) on water hexamers and observed 2-body errors of up to 10-30 kcal/mol (0.4-1.2 eV). Given MB-Pol's definition through many-body interactions, the large 2-body errors are concerning; they imply strong error cancellations that explain hindered transferability to the vapor-liquid equilibrium or the gas phase. Concerningly, even the direct inclusion of gas-phase water cluster data in the MLIP training did not fully resolve the MBD errors, and the liquid-phase properties were compromised.[30] Zhai et al.[30] therefore concluded that DeePMD does not learn the fundamental physical interactions of water and thus transferability of MLIPs to new phases is not possible.

In this work, we demonstrate that MLIPs based on equivariant graph neural networks[33] (NNs) allow for the construction of transferable water potentials. Equivariant graph NNs[34, 35] advance beyond previous invariant (and hence non-equivariant) MLIPs like DeePMD[32] by using higher-order tensors instead of scalar-only features, capturing system complexities more effectively. As a result, the constructed MLIPs also require significantly fewer training structures than Zhai et al.[30]; we require an order-of-magnitude less number of training structures to reach our goal. In addition, we show that relevant fundamental water interactions can be learned from solely liquid-phase information to quantitatively capture the many-body decomposition of gas-phase isomers, and obtain transferable potentials with accurate liquid-phase, vapor-liquid equilibrium, and solid ice properties. Finally, we show that turning off equivariance in our MLIPs undermines their ability to describe water in a transferable manner, highlighting the importance of equivariance.

We perform reference ground truth MB-Pol calculations using the MBX[19] package as implemented in Large-scale Atomic/Molecular Massively Parallel Simulator[36] (LAMMPS) to ensure agreement with prior literature. Using PackMol[37], liquid water configurations of 128 or 256 water molecules, in a 16 Å or 20 Å cube, respectively, are constructed and initially equilibrated in LAMMPS under NVT conditions at 200 K with a 0.1 fs timestep to create stable starting configurations. Production runs are performed by increasing the temperature up to the desired temperature over 5 ps with a 0.5 fs timestep, followed by an NPT simulation at the final temperature until equilibration is achieved. Vapor-liquid equilibrium (VLE) simulations of 256 water molecules are performed with the same settings, but a vacuum is added such that the Z direction is a total of 100 Å before the initial run. The final VLE simulation is kept under NVT conditions to avoid the collapse of the vacuum. Ice simulations are performed under NVT at 250 K using ice structures found in Materials Project[38, 39]. Simulation temperature and pressure are controlled via a chain of Nosé-Hoover thermostats and barostats[40, 41]. Section 9 of the Supplementary Information (SI) contains the LAMMPS input files, structures, and versions.

We train MLIPs using NequIP[40] with the Allegro[35] extension. The development branch of NequIP and Allegro downloaded on October 23, 2023, from GitHub ([https://github.com/mir-group/nequip](https://github.com/mir-group/nequip), [https://github.com/mir-group/allegro](https://github.com/mir-group/allegro)) is used for all training and inference. We choose a radial cutoff of 6.5 Å for generating neighbor lists and allow equivariant E(3) products up to $L_{max} = 2$ in the tensor layers. The neural network architecture is designed as two interaction layers: each layer includes 3 tensor product layers of 64 neurons each. The interaction layer block is followed by 3 latent layers of 64 neurons each. The feature layer comes before the interaction layer and includes 64 input features. A polynomial cutoff of 12 Å is used with a trainable Bessel basis set of 12 basis functions to form the feature descriptors. The Adam optimizer is used for training with a plateau learning rate schedule. We base the convergence of model training on a loss function, setting the energy MAE loss coefficient to 5 and the force RMSE coefficient to 1, balancing convergence to energies and forces. The stress RMSE loss coefficient is set to 100 to ensure highly accurate stresses are achieved in the resulting model. To improve the stability of the model, the MLIP is augmented by a Ziegler-Biersack-Littmark (ZBL)[42] nuclear repulsion term, preventing unphysical collapse of nuclei at small distances. Section 9 of the SI contains a sample input yaml file for training.

The water training and validation configurations are sampled from NPT molecular dynamics (MD) simulations of liquid water and contour exploration[43] (NVE$_{pot}$) MD simulations of liquid water. NPT MB-Pol MD simulations provide the initial dataset for training MLIPs. We then use the MLIP to extend the MD runs, generating additional uncorrelated training structures. We incorporate these structures back into the training set using an iterative (active) training protocol through committee selection, as described in the following paragraph. NPT MD simulations are performed in LAMMPS (as previously described) at 1 atm of pressure and a temperature range between 200 K and 368 K. Contour Exploration[43] MD simulations are performed via Atomic Simulation Environment[44] (ASE) to improve MLIP stability using an energy target of between 1 and 10 kcal/mol per water molecule at fixed volumes to explore less stable configurations. The test set of 10,000 structures is built using only NPT simulations which were equilibrated for 5 ns and sampled evenly across the last 1 ns while ensuring at least 1 ps separation between selected structures. We emphasize that our test and training/validation sets are constructed from independent MD simulations, so the MLIP training/validation sets are uncorrelated with the test set. Section 9 of the SI contains sample MD simulation input files.

To minimize data requirements and computational cost of training the MLIP, a committee[45] (otherwise known as an ensemble) of MLIP models is used to select high-error training structures in each iterative (active) learning cycle[46]. The configurations with the highest error relative to MB-Pol are included in the training dataset (up to 1000 training points in a single pass). This data selection iteratively improves the model until the test set predictions converge between iterative cycles within the MAE of 0.001 kcal/mol/atom (0.4 meV/atom), RMSE of 0.1 kcal/mol/Å (4 meV/Å), and RMSE of 0.001 kcal/Å$^2$ (0.4 meV/Å$^2$) for energy, force, and stress, respectively. Our converged MLIPs required 7 cycles with a final training set of 3200 training structures and 500 validation structures. The convergence criteria are selected to reach similar or higher final accuracy than Zhai et al.[30] We also note that our procedure is distinct from most committee models,[47, 48] which attempt to evaluate uncertainty from the standard deviation of predictions, without knowing the ground truth. In our work, we evaluate errors directly compared to the ground truth since MB-Pol is computationally affordable. We find an approximately linear correlation between the committee uncertainty and true error (Figure S1), indicating that uncertainty-based selection can also work if a more expensive method is used to generate the ground truth data. The final model energy MAE is trained to 0.0042 kcal/mol/atom (0.18 meV/atom), the force RMSE is trained to 0.64 kcal/mol/Å (27 meV/Å), and stress RMSE is trained to 0.0025 kcal/mol/Å$^2$ (0.11 meV/Å$^2$).

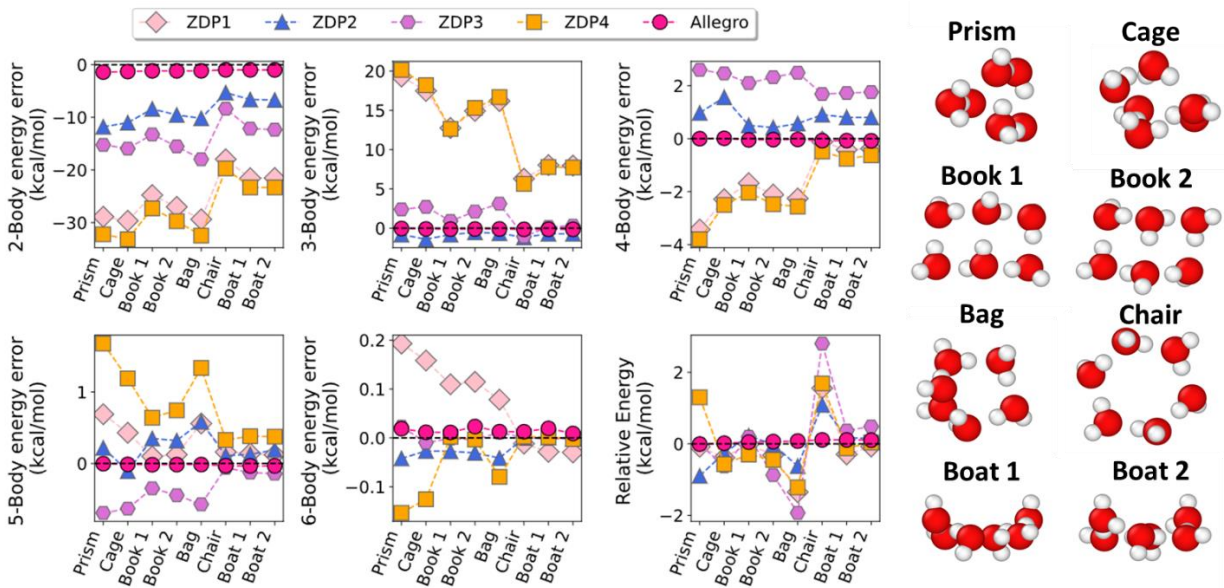

**Figure 1.** The energy errors of water hexamers and errors derived from the many-body decomposition of water into 2-6 body terms. The dashed black line at 0 kcal/mol refers to the ground truth MB-Pol result. Structures for hexamers as indicated on the x-axes are shown on the right. We emphasize that the scale on the y-axis is adjusted as needed to fit the size of the MLIP error. All DeePMD models are trained in Zhai et al.[30] and labeled as ZDPX where X is the seed number provided in their original manuscript.

We start our analysis with the many-body decomposition (MBD) of water hexamers (Figure 1) following the work of Zhai et. al.[30] to determine if fundamental interactions between water molecules are learned in our MLIP or if we also rely on significant error cancellation. We do not compare Muniz et al.[31] MLIPs here, as their work does not report on MBD. All MBD error values are in Table S2. In the MBD of water hexamers, we note a considerable error reduction, especially in the 2- and 3-body terms. The maximum 2-body error for our Allegro model is just -1.4 kcal/mol whereas the best DeePMD (ZDP2) model gives a maximum error of -11.8 kcal/mol and the worst (ZDP4) is -33.2 kcal/mol. Similarly, the 3-body error for our Allegro model is only -0.15 kcal/mol while the best DeePMD (ZDP2) model provides a maximum error of -1.4 kcal/mol, and the error for the ZDP4 model reaches 20.2 kcal/mol. The maximum errors for the 4, 5, and 6-body terms for our Allegro model are 0.07, 0.03, and 0.02 kcal/mol, respectively, which are smaller than the inherent error of our MLIP based on the test set. For this reason, we believe these errors are not relevant and most likely are just noise. For the best DeePMD (ZDP2) model, the 4, 5, and 6-body errors are instead 1.56, 0.58, and 0.04 kcal/mol respectively, and this indicates that error compensation is still occurring up to the 5-body terms. Improvements in MBD terms also result in improvements in the relative energy of each hexamer compared to the MB-Pol energy of the prism structure. The maximum error in predicting hexamer energy is only 0.11 kcal/mol with our Allegro model, while the maximum error is 1.08 kcal/mol with the best DeePMD (ZDP2) model whereas the worst DeePMD model (ZDP3) shows a maximum error of 2.8 kcal/mol. The small 0.11 kcal/mol error present in hexamer energies for our Allegro model is due to the 2-body errors being almost fully compensated with the remaining errors in the 1-6 body terms. This indicates that the Allegro model still relies on some minor error cancellations. Relative to the prism structure (Figure S3), our equivariant Allegro model's hexamer energy predictions show a mean error of just 0.05 kcal/mol, accurately predicting water hexamers' stability order, which is not qualitatively achieved by the ZDP models[30]. By comparing with the chemical accuracy (1 kcal/mol)[49] of the CC data that is used to fit MB-Pol, we can see that our Allegro MLIP can reproduce MB-Pol across all MBD terms (1-6 body) within chemical accuracy. Our Allegro model delivers a precise many-body decomposition (MBD) of water hexamers within the accuracy of CC data.

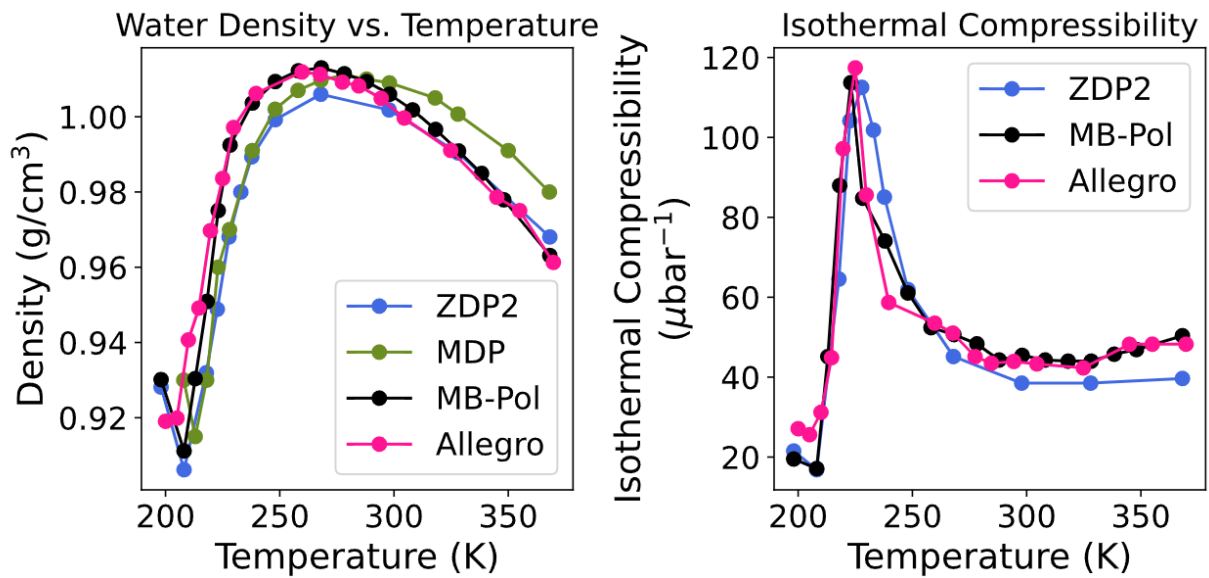

**Figure 2.** Equilibrium liquid-phase density (left) and isothermal compressibility (right) as calculated from NPT simulations of bulk liquid water. ZDP2 corresponds to the best non-augmented DeePMD model from Zhai et al.[30] and MDP corresponds to the DeePMD model from Muniz et al.[31]

While MBD offers a static analysis of the trained potential at the energy surface minima, describing dynamic structures and their ensemble average properties remains critical and we demonstrate that via MD simulations. We analyze the density and isothermal compressibility of liquid water across a large temperature range, which require an accurate and stable MLIP for the duration of the MD simulation. We compare our simulations with the best ZDP model (ZDP2), MDP model, and MB-Pol directly. Across a broad temperature span of 230 – 365 K, our Allegro MLIP demonstrates quantitative agreement, within 0.003 g/cm³, as seen in Figure 2. In contrast, DeepMD MLIPs[30, 31] exhibit errors up to 0.02 g/cm3 within the same range. Deviations are also seen in the Allegro model in the deep supercooling region below 230 K (±0.01 g/cm³). This is expected given that the long equilibration times required to obtain statistically relevant data in this temperature regime were not pursued in this work due to computational cost (see method details).

We observe that both DeePMD models seem to perform better in different regions, indicating that it is difficult to train a model that properly describes high temperature and low temperature at the same time. ZDP2 model shows errors of <0.003 g/cm$^3$ in the 195-220 K and 300-360 K regions, while the MDP model only achieves the same error metric in the 270-300 K region. We also check the density maximum that is found between 265-270 K for MB-Pol and this is also observed for our Allegro MLIP. ZDP2 might also show density maximum at the right temperature; however, it is difficult to confirm due to the large gaps between data points (20 K between points). On the other hand, the MDP clearly provides a density maximum at ~275-280 K. We think error cancellation affects temperatures inconsistently with the DeePMD MLIPs being incapable of describing all temperatures simultaneously in previous works[30, 31].

We analyze isothermal compressibility and all MLIPs appear to find the same peak position (~225 K within ±5 K) and maximum (~114 μbar ±3 μbar) for isothermal compressibility in the supercooling region. In the non-supercooling water region (>273 K), our Allegro model out-performs DeePMD as we observe close agreement with MB-Pol isothermal compressibility (±3 μbar on average for Allegro vs. ±13 μbar for DeePMD). As the isothermal compressibility is a function of fluctuations in the cell size, we believe that errors in isothermal compressibility may be related to errors in stress, a detail not addressed in earlier studies. These isothermal compressibility results may simply indicate that the stresses in our work are trained more accurately. In the supercooling region's lowest temperatures (<215 K), the Allegro model starts to overpredict isothermal compressibility by 11 μbar. This overestimation likely originates from the initial training's lack of focus on the supercooling region as discussed in the density analysis.

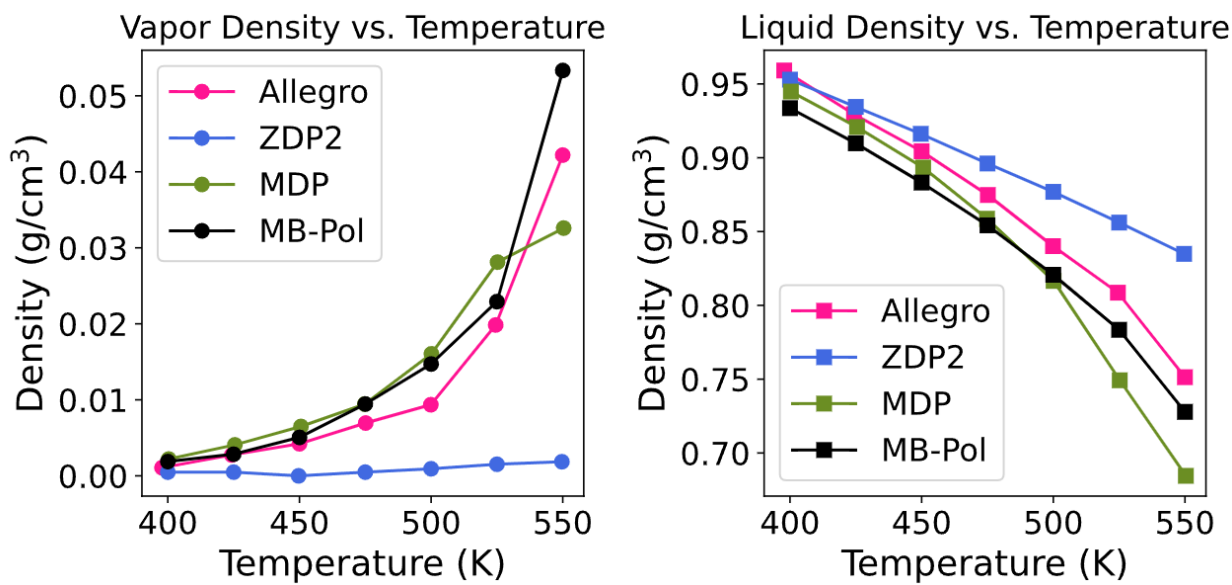

**Figure 3.** Equilibrium vapor-liquid density of the vapor phase (left) and liquid phase (right) calculated from NVT simulations of 256 water molecules. ZDP2 corresponds to the best non-augmented DeePMD model from Zhai et al.[30] and MDP corresponds to the DeePMD model from Muniz et al.[31]

We further evaluate our Allegro MLIP's capability to describe the VLE, which still contains the same fundamental interactions (covalent bonding, hydrogen bonding, dispersion interaction) that are present in the liquid phase but can be a challenge based on previous MLIP studies. Based on earlier MBD results, we also anticipate reasonable accuracy in the system's gas phase portion. Thus, this test serves as an indicator of the system's proficiency in depicting the interfacial region between liquid and gas phases. The VLE from Allegro and both DeePMD publications are shown in Figure 3 and a table of values can be found in Table S4. Simulations of the VLE using Allegro show a general improvement over previous models, especially at higher temperatures. The vapor density shows a mean absolute percent error of 11%, 91%, and 23% in the studied 400-550 K region for our Allegro, the ZDP2, and the MDP models, respectively. The vapor density error at high temperature (550 K) is lowest for Allegro (0.01 g/cm$^3$) with MDP (0.02 g/cm$^3$) and ZDP (0.05 g/cm$^3$) showing larger underpredictions. The liquid density shows a mean absolute percent error of 2.6%, 6.3%, and 2.8% in the studied 400-550 K region for our Allegro, the ZDP2, and the MDP model, respectively. At the highest temperature (550 K), the Allegro model again performs best (overprediction of 0.02 g/cm$^3$) relative to MDP (underprediction of 0.04 g/cm$^3$) and ZDP (overprediction of 0.11 g/cm$^3$).

Based on the VLE results, it appears that the original ZDP2 model without augmentation is not transferable to the VLE due to incorrect predictions of gas phase density. The lack of transferability of ZDP2 is observed in the original publication and augmentation was seen to harm the liquid phase density, so we do not consider it as a solution to transferability[30]. The MDP model performs similarly to our Allegro model for the liquid phase density in the VLE and slightly worse in the gas-phase density in the VLE but has issues in the liquid phase of water for most temperatures as previously described. Our results on VLE imply that DeePMD-trained MLIPs might lack the necessary degrees of freedom to simultaneously capture both liquid and gas phases, requiring changes to model architecture. On the contrary, our Allegro model appears to transfer well to VLE suggesting that equivariance enables training of a transferable potential without gas-phase training augmentation and that generally transferring to different phases is a challenging but solvable problem with improvements to the model architecture.

Building on our Allegro model's successful ability to transfer to VLE, we further explore its transferability to the Ih[50], Ic[51], II, and III phases of solid ice structures. We study the Ih phase as it is most common, Ic as it is also accessible at ambient pressures, and II/III as these phases are only accessible at higher pressures[52] which present a larger challenge to our liquid phase training set. We emphasize that previous ZDP and MDP models do not report on the transferability to any ice structures in their original publications, although phase diagrams have been reported for ice for ZDP in a later work[17]. We also analyze the Vibrational Density of States (VDOS) and the predictions of energy, force, and stress relative to MB-Pol for our Allegro model as seen in Figure 4 and Table S1, respectively. Here we note that comparisons with literature are not complete for the predictions of energy, force, and stress as neither of the previously reported models[30] report stress error and Muniz et al.[31] does not report energy or force error at all. For future comparisons, we recommend that MLIP reports should disclose all errors available after training.[53]

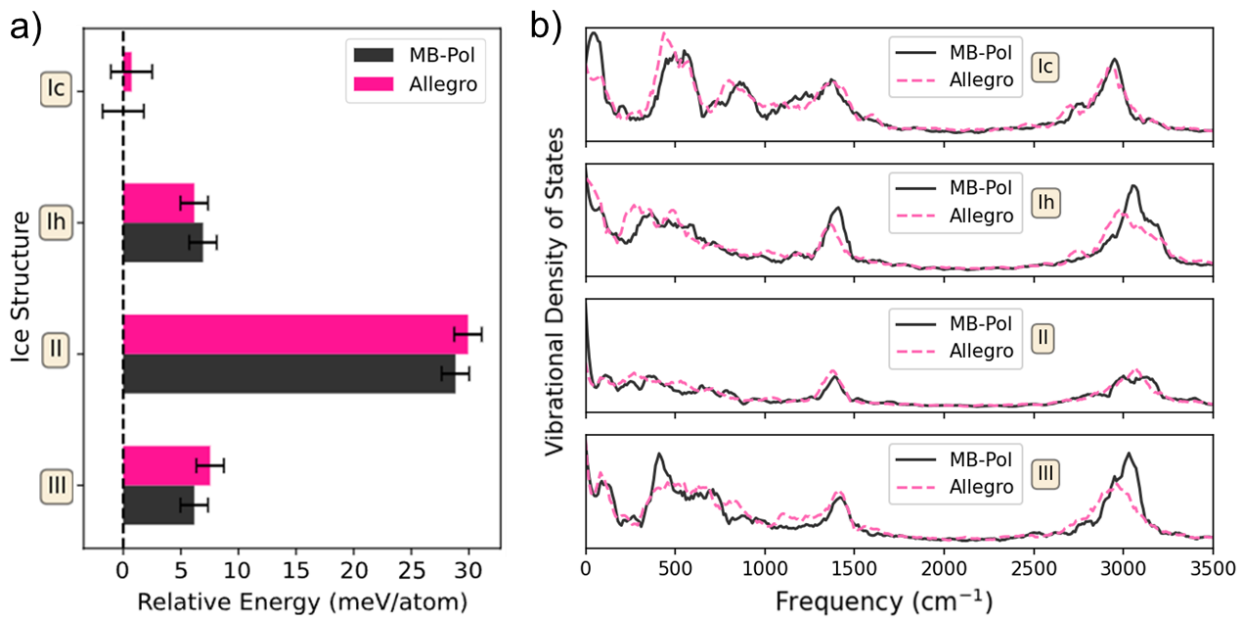

**Figure 4.** The relative energies normalized per atom (a) and vibrational density of states (b) for the tested ice phases (Ic, Ih, II, and III from top to bottom) for our Allegro model and reference MB-Pol. A description of the VDOS method / filtering can be found in Section 6 of the SI. Error bars shown on relative energies are given by a chemical accuracy uncertainty normalized by the number of atoms.

As expected, we observe greater errors for ice structures relative to the liquid water training set, but the errors are still below the errors that have been recently reported for other MLIPs in current literature. For example, errors for MLIP studies[54-56] including studies related to water[57-59] can be up to 0.11 kcal/mol/atom (5 meV/atom) and 2.2 kcal/mol/Å (100 meV/Å) for energy and force, respectively. Generally, solid ice phases have approximately 3-4 times the error in energy and force than our liquid water test set (Table S1). Stress errors are larger at between 6-8 times the liquid water test set (Table S1). To provide a clearer picture of the magnitude of energy errors, we investigate the relative energy of each ice structure (Figure 4a) at experimental volumes and compare with a chemical accuracy of 1 kcal/mol (43 meV) normalized per atom to 0.027 kcal/mol (1.2 meV/atom), which is still a larger error than what we obtain from our ice test sets. The order of relative energies are correctly predicted within the bounds of the chemical accuracy across all structures investigated, showing that the interactions present in solid ice are properly captured by our equivariant MLIP model. The errors observed may also be explained by the long-range dipole interactions present which are not captured well by a short-ranged NN such as Allegro. We hypothesize that message-passing[60] equivariant NNs such as NequIP[34] or MACE[61] may even perform better for this task as the message-passing nature can allow information propagate beyond the chosen radial cutoff. The ability to extrapolate to the ice phase in terms of error, force, and stress demonstrates the effectiveness of equivariance as the structures contain geometries which are not expected to be present in the training structures, due to their highly ordered nature.

We also analyze the VDOS and our Allegro model reproduces all the present features qualitatively (Figure 4b). The peak positions in VDOS are reproduced quantitatively after rescaling Allegro results by 1.04 indicating that the dynamics of ice are captured well with a well-defined systematic error (Figure 4). The peak intensities in the VDOS are also correct within a percent error of 30% across the range, with most regions having a percent error lower than 10%. The investigation of the solid phase highlights that our Allegro model is transferable in a way not previously reported by Zhai et al.[30] or Muniz et al.[31]

We hypothesize that equivariance, through the use of higher-order tensors as features[62], allows for higher accuracies to be achieved when compared with invariant neural networks such as DeePMD. For testing of this hypothesis, we train a non-equivariant Allegro model, incorporating only invariant scalar features ($L_{max} = 0$) akin to previous DeePMD models. We provide a direct comparison of the non-equivariant Allegro model with our equivariant Allegro model and the best ZDP2 DeePMD model via MBD analysis in Figure S2 and Table S1. The maximum 2-body MBD error of the equivariant Allegro model (-1.42 kcal/mol) increases to -7.64 kcal/mol when equivariance is removed, which is similar to the error of the invariant ZDP2 model (-11.80 kcal/mol). The maximum 3-body MBD errors of the equivariant Allegro, the non-equivariant Allegro, and the ZDP2 model correspond to -0.15 kcal/mol, 0.74 kcal/mol, and -1.41 kcal/mol, respectively, again indicating similarly high errors for the non-equivariant Allegro and DeePMD models. Another striking similarity between the non-equivariant Allegro and the ZDP2 model is the difficulty in describing the relative energy of the Bag and Chair hexamers. Figure S2 and S3 show a similar order-of-magnitude over- and under-stabilization increase in relative energy error of the Bag and Chair hexamers for the non-equivariant Allegro and the ZDP2 model, respectively. We also note that the non-equivariant Allegro model is found to be unstable during MD simulations, indicating another advantage of the inclusion of equivariance. The non-equivariant Allegro model confirms that equivariance enables NN models to be more accurate and stable with the same input training structures, and that equivariant models are more transferable due to their flexible descriptors, capturing physics more effectively.

In conclusion, we find that the inclusion of equivariant features makes MLIPs generally transferable as MLIPs trained solely on liquid water data can accurately simulate liquid-phase properties, gas-phase clusters, VLE, and even solid ice properties. Equivariant NNs show order-of-magnitude improvements in the MBD analysis of water hexamers particularly for the 2-body and 3-body terms and are within the reach of the accuracy of the underlying CC data used for developing MP-Pol. The improvements in MBD result in the prediction of the correct stability order for gas phase isomers and indicates that equivariant NNs learn more correct fundamental interactions between water molecules from liquid-phase data. Apart from gas-phase and VLE transferability, we also show surprisingly good transferability to solid ice phases, where unseen high pressure and ordered structures are predicted within chemical accuracy (1 kcal/mol) and a qualitatively correct VDOS is obtained, indicating the error of the MLIP is similar to the underlying CC method. In addition, we demonstrate that the MBD fails when equivariance is not included in our Allegro model, confirming the importance of equivariance. Our results represent an important step forward towards fast and universally transferable MLIPs for water. The presented results also suggest indirectly that transferable MLIPs might be possible to construct from CC data of gas-phase water clusters and be transferable to bulk phases[63]. Such MLIPs would circumvent the need for intermediate potentials and may yield properties and phase diagrams of water with hitherto unreached accuracy. As a closing thought, we believe equivariant NNs, such as Allegro, will widely impact simulations of biological, environmental, and electrochemical systems due to the accuracy provided by the inclusion of equivariance.

**Supporting Information**.

The following files are available free of charge.

The supplementary information is available containing additional information on methods applied, analysis results, and tabulated data for future comparisons. (PDF)

Training data is publicly available on Zenodo. Additional information and trajectories for molecular dynamics are available from the corresponding author upon request.

**Author Information**

**Corresponding Author**

**Tibor Szilvási** - Department of Chemical and Biological Engineering, University of Alabama, Tuscaloosa, AL 35487, United States; Email: tibor.szilvasi@ua.edu

**Authors**

**Tristan Maxson** - Department of Chemical and Biological Engineering, University of Alabama, Tuscaloosa, AL 35487, United States

**Author Contributions**

The manuscript was written through contributions of all authors. All authors have given approval to the final version of the manuscript. All authors have contributed equally.

**Notes**

The authors declare no competing financial interest.

**Acknowledgements**

T.M. and T.S. would like to acknowledge the financial support of the Department of Energy Basic Energy Sciences (DOE-BES) under grant number DE-SC0024654. The authors thank Sophia Ezendu, Gbolagade Olajide, and Ademola Soyemi for their insightful comments on the manuscript. T.M and T.S. would also like to thank the University of Alabama and the Office of Information Technology for providing high-performance computing resources and support that has contributed to these research results. This work was also made possible in part by a grant of high-performance computing resources and technical support from the Alabama Supercomputer Authority. This material is based upon work supported by the U.S. Department of Energy, Office of Science, Office of Advanced Scientific Computing Research, Department of Energy Computational Science Graduate Fellowship under Award Number(s) DE-SC0023112. This research used resources of the National Energy Research Scientific Computing Center (NERSC), a U.S. Department of Energy Office of Science User Facility located at Lawrence Berkeley National Laboratory, operated under Contract No. DE-AC02-05CH11231 using NERSC award BES-ERCAP0024218. This work was also made possible in part by the National Science Foundation under grant number 2245120. Any opinions, findings, conclusions, and/or recommendations expressed in this material are those of the authors(s) and do not necessarily reflect the views of funding agencies.

**References**

(1) Zhao, S.; Ramirez, R.; Vuilleumier, R.; Borgis, D. Molecular density functional theory of solvation: From polar solvents to water. *The Journal of chemical physics* **2011**, *134* (19).

(2) Heenen, H. H.; Gauthier, J. A.; Kristoffersen, H. H.; Ludwig, T.; Chan, K. Solvation at metal/water interfaces: An ab initio molecular dynamics benchmark of common computational approaches. *The Journal of Chemical Physics* **2020**, *152* (14).

(3) Steinmann, S. N.; Sautet, P.; Michel, C. Solvation free energies for periodic surfaces: comparison of implicit and explicit solvation models. *Physical Chemistry Chemical Physics* **2016**, *18* (46), 31850-31861.

(4) Benjamin, I. Chemical reactions and solvation at liquid interfaces: A microscopic perspective. *Chemical reviews* **1996**, *96* (4), 1449-1476.

(5) Cooke, R.; Kuntz, I. The properties of water in biological systems. *Annual review of biophysics and bioengineering* **1974**, *3* (1), 95-126.

(6) Hosseinpour, S.; Tang, F.; Wang, F.; Livingstone, R. A.; Schlegel, S. J.; Ohto, T.; Bonn, M.; Nagata, Y.; Backus, E. H. Chemisorbed and physisorbed water at the TiO2/water interface. *The journal of physical chemistry letters* **2017**, *8* (10), 2195-2199.

(7) Mark, P.; Nilsson, L. Structure and Dynamics of the TIP3P, SPC, and SPC/E Water Models at 298 K. *The Journal of Physical Chemistry A* **2001**, *105* (43), 9954-9960. DOI: 10.1021/jp003020w.

(8) Mizan, T. I.; Savage, P. E.; Ziff, R. M. Molecular dynamics of supercritical water using a flexible SPC model. *The Journal of Physical Chemistry* **1994**, *98* (49), 13067-13076.

(9) Hu, H.; Ma, Z.; Wang, F. On the transferability of three water models developed by adaptive force matching. In *Annual Reports in Computational Chemistry*, Vol. 10; Elsevier, 2014; pp 25-43.

(10) Leven, I.; Hao, H.; Tan, S.; Guan, X.; Penrod, K. A.; Akbarian, D.; Evangelisti, B.; Hossain, M. J.; Islam, M. M.; Koski, J. P. Recent advances for improving the accuracy, transferability, and efficiency of reactive force fields. *Journal of chemical theory and computation* **2021**, *17* (6), 3237-3251.

(11) Glättli, A.; Oostenbrink, C.; Daura, X.; Geerke, D. P.; Yu, H.; Van Gunsteren, W. F. On the transferability of the SPC/L water model to biomolecular simulation. *Brazilian journal of physics* **2004**, *34*, 116-125.

(12) Babin, V.; Leforestier, C.; Paesani, F. Development of a "First Principles" Water Potential with Flexible Monomers: Dimer Potential Energy Surface, VRT Spectrum, and Second Virial Coefficient. *Journal of Chemical Theory and Computation* **2013**, *9* (12), 5395-5403. DOI: 10.1021/ct400863t.

(13) Babin, V.; Medders, G. R.; Paesani, F. Development of a "First Principles" Water Potential with Flexible Monomers. II: Trimer Potential Energy Surface, Third Virial Coefficient, and Small Clusters. *Journal of Chemical Theory and Computation* **2014**, *10* (4), 1599-1607. DOI: 10.1021/ct500079y.

(14) Medders, G. R.; Babin, V.; Paesani, F. Development of a "First-Principles" Water Potential with Flexible Monomers. III. Liquid Phase Properties. *Journal of Chemical Theory and Computation* **2014**, *10* (8), 2906-2910. DOI: 10.1021/ct5004115.

(15) Muniz, M. C.; Gartner, T. E., III; Riera, M.; Knight, C.; Yue, S.; Paesani, F.; Panagiotopoulos, A. Z. Vapor–liquid equilibrium of water with the MB-pol many-body potential. *The Journal of Chemical Physics* **2021**, *154* (21), 211103. (acccessed 2/3/2024).

(16) Zhu, X.; Riera, M.; Bull-Vulpe, E. F.; Paesani, F. MB-pol(2023): Sub-chemical Accuracy for Water Simulations from the Gas to the Liquid Phase. *Journal of Chemical Theory and Computation* **2023**, *19* (12), 3551-3566. DOI: 10.1021/acs.jctc.3c00326.

(17) Bore, S. L.; Paesani, F. Realistic phase diagram of water from "first principles" data-driven quantum simulations. *Nature Communications* **2023**, *14* (1), 3349.

(18) Reddy, S. K.; Straight, S. C.; Bajaj, P.; Huy Pham, C.; Riera, M.; Moberg, D. R.; Morales, M. A.; Knight, C.; Götz, A. W.; Paesani, F. On the accuracy of the MB-pol many-body potential for water: Interaction energies, vibrational frequencies, and classical thermodynamic and dynamical properties from clusters to liquid water and ice. *The Journal of chemical physics* **2016**, *145* (19).

(19) Reddy, S. K.; Moberg, D. R.; Straight, S. C.; Paesani, F. Temperature-dependent vibrational spectra and structure of liquid water from classical and quantum simulations with the MB-pol potential energy function. *The Journal of Chemical Physics* **2017**, *147* (24).

(20) Pham, C. H.; Reddy, S. K.; Chen, K.; Knight, C.; Paesani, F. Many-body interactions in ice. *Journal of Chemical Theory and Computation* **2017**, *13* (4), 1778-1784.

(21) Gartner III, T. E.; Hunter, K. M.; Lambros, E.; Caruso, A.; Riera, M.; Medders, G. R.; Panagiotopoulos, A. Z.; Debenedetti, P. G.; Paesani, F. Anomalies and local structure of liquid water from boiling to the supercooled regime as predicted by the many-body MB-pol model. *The Journal of Physical Chemistry Letters* **2022**, *13* (16), 3652-3658.

(22) Deringer, V. L.; Caro, M. A.; Csányi, G. Machine learning interatomic potentials as emerging tools for materials science. *Advanced Materials* **2019**, *31* (46), 1902765.

(23) Zuo, Y.; Chen, C.; Li, X.; Deng, Z.; Chen, Y.; Behler, J. r.; Csányi, G.; Shapeev, A. V.; Thompson, A. P.; Wood, M. A. Performance and cost assessment of machine learning interatomic potentials. *The Journal of Physical Chemistry A* **2020**, *124* (4), 731-745.

(24) Mishin, Y. Machine-learning interatomic potentials for materials science. *Acta Materialia* **2021**, *214*, 116980.

(25) Mueller, T.; Hernandez, A.; Wang, C. Machine learning for interatomic potential models. *The Journal of chemical physics* **2020**, *152* (5).

(26) Ye, H.-f.; Wang, J.; Zheng, Y.-g.; Zhang, H.-w.; Chen, Z. Machine learning for reparameterization of four-site water models: TIP4P-BG and TIP4P-BGT. *Physical Chemistry Chemical Physics* **2021**, *23* (17), 10164-10173.

(27) Yu, Q.; Qu, C.; Houston, P. L.; Nandi, A.; Pandey, P.; Conte, R.; Bowman, J. M. A status report on "Gold Standard" machine-learned potentials for water. *The Journal of Physical Chemistry Letters* **2023**, *14* (36), 8077-8087.

(28) Paesani, F. Getting the right answers for the right reasons: Toward predictive molecular simulations of water with many-body potential energy functions. *Accounts of Chemical Research* **2016**, *49* (9), 1844-1851.

(29) Cisneros, G. A.; Wikfeldt, K. T.; Ojamäe, L.; Lu, J.; Xu, Y.; Torabifard, H.; Bartók, A. P.; Csányi, G.; Molinero, V.; Paesani, F. Modeling molecular interactions in water: From pairwise to many-body potential energy functions. *Chemical reviews* **2016**, *116* (13), 7501-7528.

(30) Zhai, Y.; Caruso, A.; Bore, S. L.; Luo, Z.; Paesani, F. A "short blanket" dilemma for a state-of-the-art neural network potential for water: Reproducing experimental properties or the physics of the underlying many-body interactions? *The Journal of Chemical Physics* **2023**, *158* (8), 084111.

(31) Muniz, M. C.; Car, R.; Panagiotopoulos, A. Z. Neural Network Water Model Based on the MB-Pol Many-Body Potential. *The Journal of Physical Chemistry B* **2023**, *127* (42), 9165-9171.

(32) Wang, H.; Zhang, L.; Han, J.; Weinan, E. DeePMD-kit: A deep learning package for many-body potential energy representation and molecular dynamics. *Computer Physics Communications* **2018**, *228*, 178-184.

(33) Geiger, M.; Smidt, T. e3nn: Euclidean neural networks. *arXiv preprint arXiv:2207.09453* **2022**. (acccessed 2/19/24).

(34) Batzner, S.; Musaelian, A.; Sun, L.; Geiger, M.; Mailoa, J. P.; Kornbluth, M.; Molinari, N.; Smidt, T. E.; Kozinsky, B. E(3)-equivariant graph neural networks for data-efficient and accurate interatomic potentials. *Nature Communications* **2022**, *13* (1), 2453. DOI: 10.1038/s41467-022-29939-5.

(35) Musaelian, A.; Batzner, S.; Johansson, A.; Sun, L.; Owen, C. J.; Kornbluth, M.; Kozinsky, B. Learning local equivariant representations for large-scale atomistic dynamics. *Nature Communications* **2023**, *14* (1), 579.

(36) Thompson, A. P.; Aktulga, H. M.; Berger, R.; Bolintineanu, D. S.; Brown, W. M.; Crozier, P. S.; in 't Veld, P. J.; Kohlmeyer, A.; Moore, S. G.; Nguyen, T. D.; et al. LAMMPS - a flexible simulation tool for particle-based materials modeling at the atomic, meso, and continuum scales. *Computer Physics Communications* **2022**, *271*, 108171.

(37) Martínez, L.; Andrade, R.; Birgin, E. G.; Martínez, J. M. PACKMOL: A package for building initial configurations for molecular dynamics simulations. *Journal of computational chemistry* **2009**, *30* (13), 2157-2164.

(38) Jain, A.; Ong, S. P.; Hautier, G.; Chen, W.; Richards, W. D.; Dacek, S.; Cholia, S.; Gunter, D.; Skinner, D.; Ceder, G.; et al. Commentary: The Materials Project: A materials genome approach to accelerating materials innovation. *APL Materials* **2013**, *1* (1), 011002. DOI: 10.1063/1.4812323 (acccessed 2/3/2024).

(39) Ong, S. P.; Richards, W. D.; Jain, A.; Hautier, G.; Kocher, M.; Cholia, S.; Gunter, D.; Chevrier, V. L.; Persson, K. A.; Ceder, G. Python Materials Genomics (pymatgen): A robust, open-source python library for materials analysis. *Computational Materials Science* **2013**, *68*, 314-319.

(40) Evans, D. J.; Holian, B. L. The nose–hoover thermostat. *The Journal of chemical physics* **1985**, *83* (8), 4069-4074.

(41) Martyna, G. J.; Klein, M. L.; Tuckerman, M. Nosé–Hoover chains: The canonical ensemble via continuous dynamics. *The Journal of chemical physics* **1992**, *97* (4), 2635-2643.

(42) Ziegler, J. F.; Biersack, J. P. The Stopping and Range of Ions in Matter. In *Treatise on Heavy-Ion Science: Volume 6: Astrophysics, Chemistry, and Condensed Matter*, Bromley, D. A. Ed.; Springer US, 1985; pp 93-129.

(43) Waters, M. J.; Rondinelli, J. M. Energy contour exploration with potentiostatic kinematics. *Journal of Physics: Condensed Matter* **2021**, *33* (44), 445901.

(44) Hjorth Larsen, A.; Jørgen Mortensen, J.; Blomqvist, J.; Castelli, I. E.; Christensen, R.; Dułak, M.; Friis, J.; Groves, M. N.; Hammer, B.; Hargus, C.; et al. The atomic simulation environment—a Python library for working with atoms. *Journal of Physics: Condensed Matter* **2017**, *29* (27), 273002.

(45) Jeong, W.; Yoo, D.; Lee, K.; Jung, J.; Han, S. Efficient atomic-resolution uncertainty estimation for neural network potentials using a replica ensemble. *The Journal of Physical Chemistry Letters* **2020**, *11* (15), 6090-6096.

(46) Wilson, N.; Willhelm, D.; Qian, X.; Arróyave, R.; Qian, X. Batch active learning for accelerating the development of interatomic potentials. *Computational Materials Science* **2022**, *208*, 111330.

(47) Artrith, N.; Behler, J. High-dimensional neural network potentials for metal surfaces: A prototype study for copper. *Physical Review B* **2012**, *85* (4), 045439.

(48) Zhu, A.; Batzner, S.; Musaelian, A.; Kozinsky, B. Fast uncertainty estimates in deep learning interatomic potentials. *The Journal of Chemical Physics* **2023**, *158* (16).

(49) Helgaker, T.; Ruden, T. A.; Jørgensen, P.; Olsen, J.; Klopper, W. A priori calculation of molecular properties to chemical accuracy. *Journal of Physical Organic Chemistry* **2004**, *17* (11), 913-933.

(50) Kuhs, W. F.; Lehmann, M. S. The Structure of Ice-Ih. In *Water Science Reviews 2: Crystalline Hydrates*, Franks, F. Ed.; Water Science Review, Vol. 2; Cambridge University Press, 1986; pp 1-66.

(51) Hagen, W.; Tielens, A. G. G. M.; Greenberg, J. M. The infrared spectra of amorphous solid water and ice Ic between 10 and 140 K. *Chemical Physics* **1981**, *56* (3), 367-379.

(52) Sanz, E.; Vega, C.; Abascal, J. L. F.; MacDowell, L. G. Phase Diagram of Water from Computer Simulation. *Physical Review Letters* **2004**, *92* (25), 255701. DOI: 10.1103/PhysRevLett.92.255701.

(53) Maxson, T.; Soyemi, A.; Chen, B. W. J.; Szilv'asi, T. Enhancing the Quality and Reliability of Machine Learning Interatomic Potentials through Better Reporting Practices. *arXiv* **2024**, (2401.02284). (acccessed 2/19/24).

(54) Goeminne, R.; Vanduyfhuys, L.; Van Speybroeck, V.; Verstraelen, T. DFT-Quality Adsorption Simulations in Metal–Organic Frameworks Enabled by Machine Learning Potentials. *Journal of Chemical Theory and Computation* **2023**, *19* (18), 6313-6325. DOI: 10.1021/acs.jctc.3c00495.

(55) Rice, P. S.; Liu, Z.-P.; Hu, P. Hydrogen Coupling on Platinum Using Artificial Neural Network Potentials and DFT. *The Journal of Physical Chemistry Letters* **2021**, *12* (43), 10637-10645. DOI: 10.1021/acs.jpclett.1c02998.

(56) Waters, M. J.; Rondinelli, J. M. Benchmarking structural evolution methods for training of machine learned interatomic potentials. *Journal of Physics: Condensed Matter* **2022**, *34* (38), 385901. DOI: 10.1088/1361-648X/ac7f73.

(57) Piaggi, Pablo M.; Selloni, A.; Panagiotopoulos, A. Z.; Car, R.; Debenedetti, P. G. A first-principles machine-learning force field for heterogeneous ice nucleation on microcline feldspar. *Faraday Discussions* **2024**, 10.1039/D3FD00100H. DOI: 10.1039/D3FD00100H.

(58) Zheng, M.; Bukowski, B. C. Probing the role of acid site distribution on water structure in aluminosilicate zeolites: insights from molecular dynamics. *ChemRxiv* **2023**.

(59) Li, L.; Calegari Andrade, M. F.; Car, R.; Selloni, A.; Carter, E. A. Characterizing Structure-Dependent TiS2/Water Interfaces Using Deep-Neural-Network-Assisted Molecular Dynamics. *The Journal of Physical Chemistry C* **2023**, *127* (20), 9750-9758. DOI: 10.1021/acs.jpcc.2c08581.

(60) Gilmer, J.; Schoenholz, S. S.; Riley, P. F.; Vinyals, O.; Dahl, G. E. Neural Message Passing for Quantum Chemistry. In Proceedings of the 34th International Conference on Machine Learning, Proceedings of Machine Learning Research; 2017.

(61) Batatia, I.; Kov'acs, D. a. P. e.; Simm, G. N. C.; Ortner, C.; Csányi, G. MACE: Higher Order Equivariant Message Passing Neural Networks for Fast and Accurate Force Fields. *ArXiv* **2022**, *abs/2206.07697*. (acccessed 2/19/24).

(62) Gao, H.; Wang, J.; Sun, J. Improve the performance of machine-learning potentials by optimizing descriptors. *The Journal of chemical physics* **2019**, *150 24*, 244110.

(63) Schran, C.; Behler, J. r.; Marx, D. Automated fitting of neural network potentials at coupled cluster accuracy: Protonated water clusters as testing ground. *Journal of chemical theory and computation* **2019**, *16* (1), 88-99.

# Transferable Water Potentials Using Equivariant Neural Networks

Tristan Maxson. Tibor Szilvási[*]

[*]Corresponding author. Email: tibor.szilvasi@ua.edu

Department of Chemical and Biological Engineering,
University of Alabama, Tuscaloosa, AL 35487, United States

## Table of Contents

## 1.    Data Digitalization

Literature data has been digitalized using WebPlotDigitizer[1] from Zhai et. al[2] and Muniz et. al[3] for comparisons to our models when raw tabulated data was not available. All data from Zhai et al. [2] was digitalized from the published manuscript. Muniz et al.[3] supplied some tabulated data and digitalization was used to confirm correctness of the table, with the final values being taken from the table when available. We refer readers to the original publications for the original datasets. Digitalization is performed by a non-rotationally calibrated manual point-by-point extraction via the web interface using the highest resolution version of the figure available. Digitalization error is expected to be <1% in the x and y axes of the plots and small deviations are not expected to change any conclusions in this manuscript.

## 2.    Comparison of Uncertainty and Error via Committee

We provide a comparison of uncertainty (standard deviation of model predictions) and ground-truth error via the committee of 3 models used for data selection in Figure S1. In this work, we chose to use the error with regard to MB-Pol rather than uncertainty for data selection because of the relatively low computational cost of the MB-Pol potential. This comparison is useful as it is common in literature for committees[4, 5] to use an uncertainty measure (such as standard deviation of predictions) when it is too computationally expensive to evaluate errors for all potential training structures due to the cost of DFT/CC calculations. Our results in Figure S1 show that an approximately linear correlation exists between error and uncertainty ($R^2 = 0.921$) indicating that the same data selection can be possible using uncertainty instead of error. So, we think our MLIP training methodology will also work for an MLIP trained on DFT calculations. Alternative approaches would be to use an uncertainty derived directly from the model architecture[6].

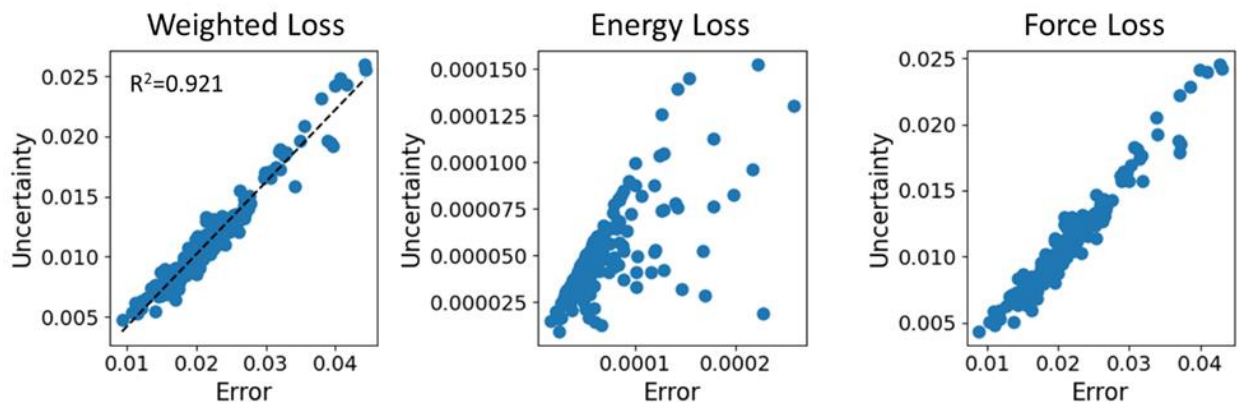

**Figure S1.** The uncertainty (standard deviation of model predictions) and ground-truth error with regard to MB-Pol are presented multiplied by their loss coefficients for the weighted loss (left), energy loss (middle), and force loss (right) at the end of the committee training. The weighted loss is weighted with the energy loss as 10 times the force loss. In this comparison, the losses are based on the mean absolute error of energy and root mean-square error of force. A linear fit with an $R^2 = 0.921$ is shown to highlight the relationship of the weighted loss error and uncertainty as a black dashed line in the left plot.

## 3. Many-Body Decomposition Method

The many-body decomposition is performed by calculating the energy error terms from the ground up, starting with 1-body error which is defined as the difference in energy between MB-Pol and the tested model for single molecules in the geometry as-is in the hexamer structure. The 2-body error is then defined as the difference in energy between MB-Pol and the tested model for pairs of molecules in the geometry as-is in the hexamer structure, but with the 1-body error subtracted. The higher body errors are defined similarly in a recursive manner, represented in Equation 1. $\varepsilon_{MBD}^N$ represents the MBD error of N bodies for a given hexamer. $(E_{MLIP} - E_{MB-Pol})$ represents the energy difference between the MLIP and MB-Pol for the partial isomer. $\sum_{M=1}^{N-1} \varepsilon_{MBD}^M$ is the summation of all lower body MBD errors for the hexamer. As a result, the error of an N-body term is only the error of that interaction and neglects all lower body error terms.

$$\varepsilon_{MBD}^N = (E_{MLIP} - E_{MB-Pol}) - \sum_{M=1}^{N-1} \varepsilon_{MBD}^M \qquad (1)$$

## 4. Many-Body Decomposition: The Role of Equivariance

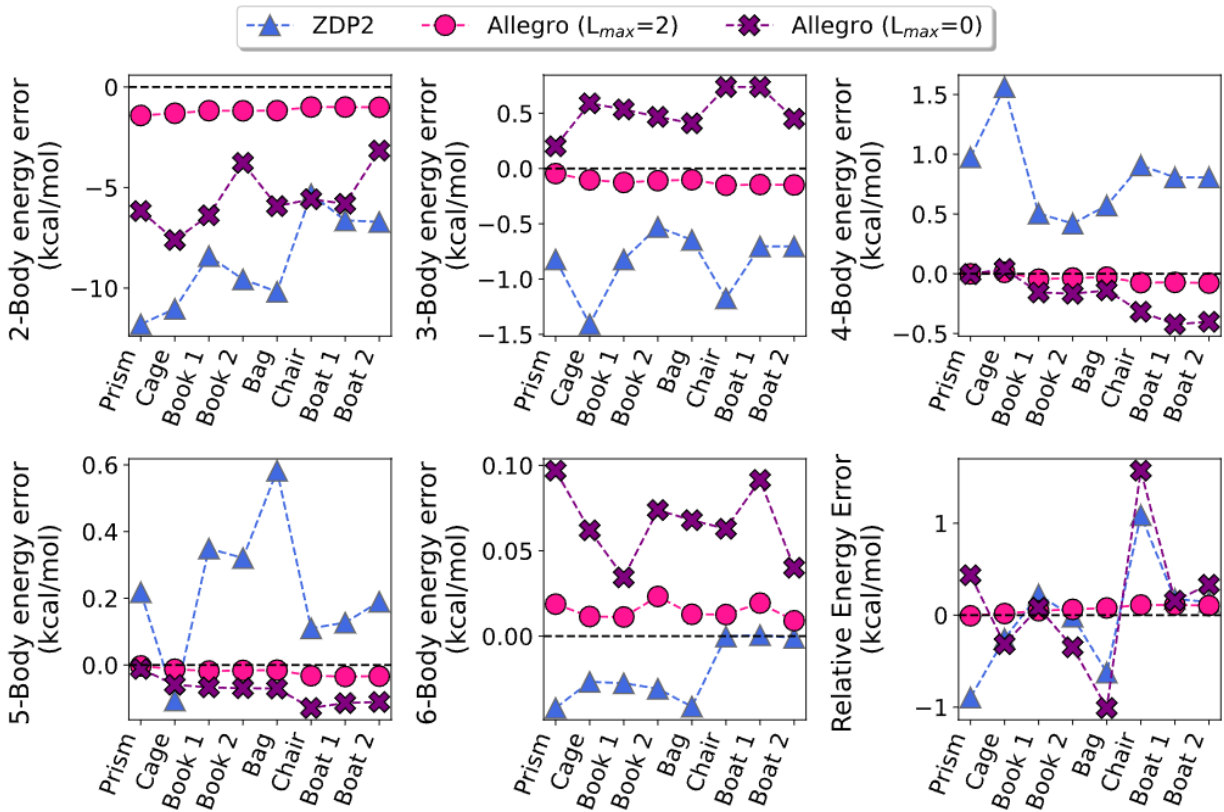

**Figure S2.** The energy errors of water hexamers and errors derived from the many-body decomposition of water into 2-6 body terms. The dashed black line at 0 kcal/mol refers to the ground truth MB-Pol result. Structures for hexamers can be found in Figure 1. We emphasize that the scale on the y-axis is adjusted as needed to fit the size of the MLIP error. All DeePMD models are trained in Zhai et al.[2] and labeled as ZDPX where X is the seed number provided in their original manuscript.

## 5. Hexamer Energies

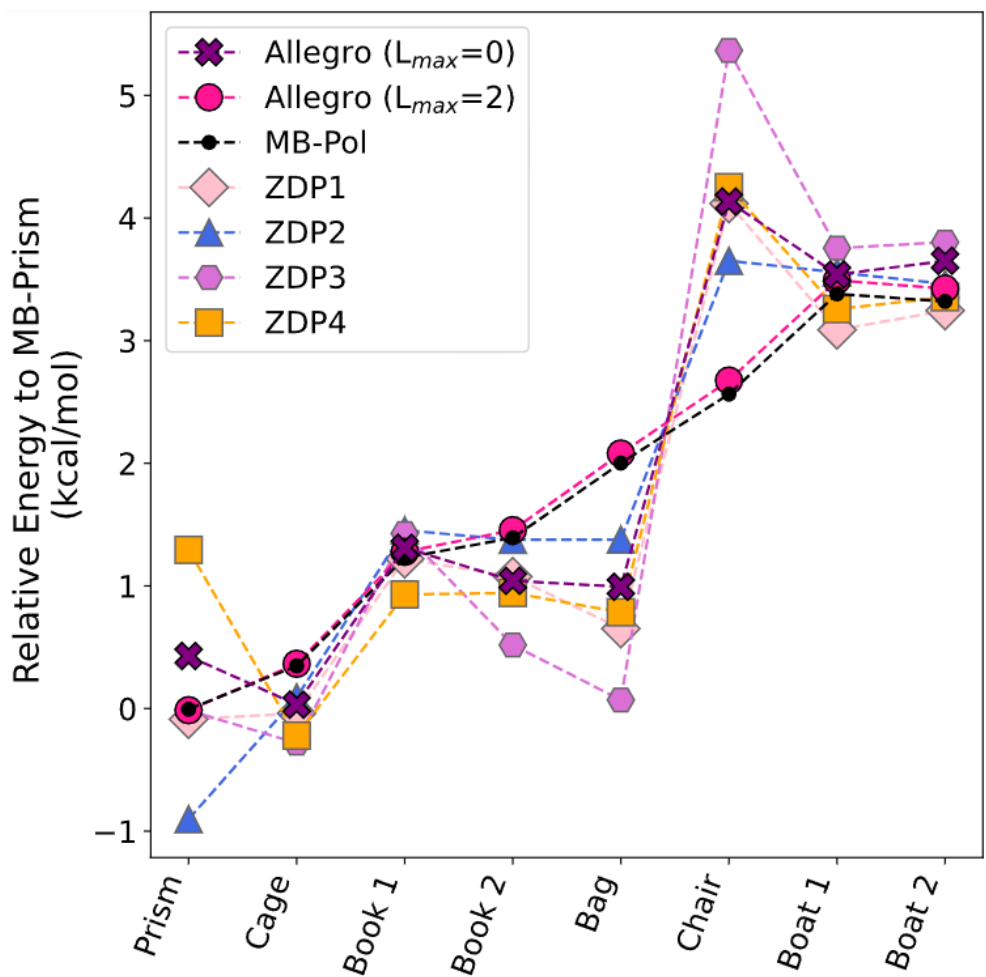

**Figure S3.** The energies of water hexamers relative to the MB-Pol energy of the prism structure. Structures can be found in Figure 1. All DeePMD models are trained in Zhai et. al[2] and are labeled as ZDPX where X is the seed number provided in their original manuscript.

## 6.    Ice Vibrational Density of States

The vibrational density of states (VDOS) for ice are calculated from the velocity autocorrelation function, followed by a Fourier transformation to the raw VDOS, consistent with other studies of VDOS for water[7-9]. The VDOS is binned within every 10 cm$^{-1}$ from 0 to 3500 cm$^{-1}$. These datapoints are then filtered according to a Savgol filter of width 7 and polynomial order 3 to smooth the curve and eliminate noise present. Finally, we scale the frequencies by 1.04 to minimize error to correct for systematic error caused by noise in the MLIP. The VDOS is calculated by sampling the last 0.5 ns of a 2.0 ns long 0.5 fs timestep NVT simulation giving a total of 1,000,000 correlated structures in VDOS. Reduction of the timestep and sampling to 0.2 fs does not provide qualitatively different results. We provide the raw VDOS without filtering or scaling in Figure S4.

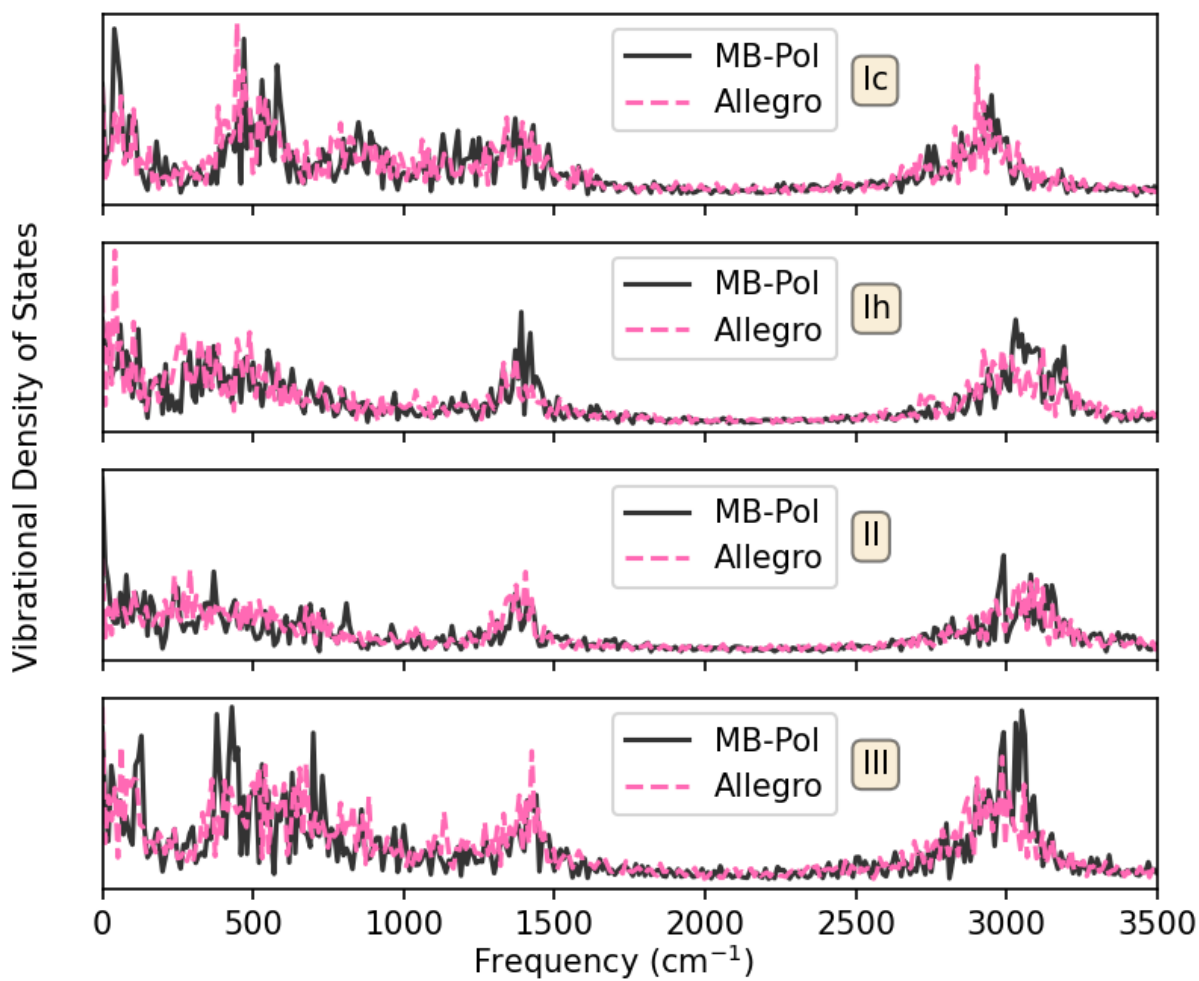

**Figure S4.** The vibrational density of states (VDOS) for the tested ice phases (Ic, Ih, II, and III from top to bottom) for the Allegro model and MB-Pol. The VDOS pictured here is not filtered or rescaled.

# 7. Statistical Analysis of Energy, Force, and Stress

**Table S1.** Statistical analysis of trained energy, force, and stress errors across ice phases and the liquid phase test sets of the Allegro model. For comparison, we also show available liquid phase test set results for ZDP2 and MDP MLIPs. The top values in the table are in units of kcal/mol, while the values in square brackets are provided in units of meV for ease of comparison.

| Phase | Energy MAE (kcal/mol/atom) [ meV/atom ] | Force RMSE (kcal/mol/Å) [ meV/Å ] | Stress RMSE (kcal/mol/Å$^2$) [ meV/Å$^2$ ] |
|---|---|---|---|
| Liquid (Allegro) | 0.0042 [ 0.18 ] | 0.64 [ 27 ] | 0.0025 [ 0.11 ] |
| Liquid (ZDP2) | 0.01 [ 0.43 ] | 1 [ 43 ] | *Not Reported* |
| Liquid (MDP) | *Not Reported* | *Not Reported* | *Not Reported* |
| Ice Ih | 0.0175 [ 0.75 ] | 2.18 [ 94 ] | 0.0424 [ 1.8 ] |
| Ice II | 0.0251 [ 1.1 ] | 2.21 [ 96 ] | 0.0280 [ 1.2 ] |
| Ice III | 0.0318 [ 1.4 ] | 1.81 [ 78 ] | 0.0152 [ 0.65 ] |
| Ice Ic | 0.0168 [ 0.72 ] | 1.99 [ 86 ] | 0.0110 [ 0.47 ] |

# 8. Tabulated Manuscript Values

## 8.1. Figure 1: Many-Body Decomposition

**Table S2.** Tabulated data for the Allegro models presented for the many-body decomposition for Figure S2 and Figure 1. Lmax=0 is the non-equivariant Allegro model found in this SI and Lmax=2 is the equivariant Allegro model presented in the main manuscript. Terms 1-6 are the many-body error term and Interact is the interaction energy relative to the MB-Pol Prism hexamer.

| Label | Term | Prism | Cage | Book 1 | Book 2 | Bag | Chair | Boat 1 | Boat 2 |
|-------|------|-------|------|--------|--------|-----|-------|--------|--------|
| Lmax=2 | 1 | 0.48 | 0.33 | 0.67 | 0.66 | 0.45 | 0.44 | 0.43 | 0.39 |
| Lmax=2 | 2 | -1.42 | -1.31 | -1.18 | -1.18 | -1.17 | -1.00 | -1.00 | -1.01 |
| Lmax=2 | 3 | -0.05 | -0.10 | -0.13 | -0.11 | -0.10 | -0.15 | -0.15 | -0.15 |
| Lmax=2 | 4 | 0.00 | 0.01 | -0.05 | -0.04 | -0.03 | -0.07 | -0.07 | -0.08 |
| Lmax=2 | 5 | 0.00 | -0.01 | -0.02 | -0.02 | -0.02 | -0.03 | -0.03 | -0.03 |
| Lmax=2 | 6 | 0.02 | 0.01 | 0.01 | 0.02 | 0.01 | 0.01 | 0.02 | 0.01 |
| Lmax=2 | Interact | -0.01 | 0.36 | 1.28 | 1.45 | 2.08 | 2.67 | 3.49 | 3.42 |
| Lmax=0 | 1 | 24.94 | 24.18 | 32.12 | 21.15 | 32.50 | 27.65 | 20.48 | 19.93 |
| Lmax=0 | 2 | -6.16 | -7.61 | -6.38 | -3.77 | -5.93 | -5.58 | -5.81 | -3.16 |
| Lmax=0 | 3 | 0.20 | 0.59 | 0.53 | 0.47 | 0.41 | 0.74 | 0.74 | 0.45 |
| Lmax=0 | 4 | 0.00 | 0.04 | -0.16 | -0.17 | -0.14 | -0.32 | -0.42 | -0.40 |
| Lmax=0 | 5 | -0.01 | -0.06 | -0.07 | -0.07 | -0.07 | -0.13 | -0.11 | -0.11 |
| Lmax=0 | 6 | 0.10 | 0.06 | 0.03 | 0.07 | 0.07 | 0.06 | 0.09 | 0.04 |
| Lmax=0 | Interact | 0.43 | 0.03 | 1.31 | 1.04 | 0.99 | 4.13 | 3.54 | 3.65 |

## 8.2. Figure 2: Liquid Phase

**Table S3.** Tabulated data for the equivariant Allegro model in Figure 2.

| Allegro Temperature (K) | Allegro Density (g/cm$^3$) | Allegro Isothermal Compressibility (μbar) |
|---|---|---|
| 200 | 0.919 | 27.100 |
| 205 | 0.920 | 25.600 |
| 210 | 0.941 | 31.200 |
| 215 | 0.949 | 44.871 |
| 220 | 0.970 | 97.200 |
| 225 | 0.984 | 117.400 |
| 230 | 0.997 | 85.610 |
| 240 | 1.006 | 58.681 |
| 260 | 1.012 | 53.510 |
| 268 | 1.011 | 51.031 |
| 277 | 1.009 | 45.132 |
| 285 | 1.008 | 43.486 |
| 294 | 1.005 | 43.923 |
| 304 | 1.000 | 43.308 |
| 325 | 0.991 | 42.314 |
| 345 | 0.979 | 48.235 |
| 355 | 0.975 | 48.244 |
| 370 | 0.961 | 48.247 |

### 8.3.    Figure 3: Vapor-Liquid Equilibrium

**Table S4.** Tabulated data for the equivariant Allegro model in Figure 3.

| Allegro Temperature (K) | Allegro Vapor Density (g/cm$^3$) | Allegro Liquid Density (g/cm$^3$) |
|---|---|---|
| 400 | 0.001 | 0.959 |
| 425 | 0.003 | 0.929 |
| 450 | 0.004 | 0.904 |
| 475 | 0.007 | 0.875 |
| 500 | 0.009 | 0.840 |
| 525 | 0.020 | 0.809 |
| 550 | 0.042 | 0.751 |

## 9.    Input Files

### 9.1.    MB-Pol: mbx.json

```
{
   "Note" :
"https://github.com/paesanilab/MBX/blob/master/examples/PEFs/001_mbpol/JSON_files/mbx_p
bc.json",
   "MBX" : {
      "twobody_cutoff"   : 9.0,
      "threebody_cutoff" : 4.5,
      "dipole_tolerance" : 1E-8,
      "dipole_method"    : "cg",
      "alpha_ewald_elec" : 0.6,
      "grid_density_elec" : 2.5,
      "spline_order_elec" : 6,
      "alpha_ewald_disp" : 0.6,
      "grid_density_disp" : 2.5,
      "spline_order_disp" : 6,
      "ff_mons"   : [],
      "connectivity_file" : "",
      "ttm_pairs" : [],
      "ignore_dispersion" : [],
      "use_lennard_jones" : [],
      "nonbonded_file" : "",
      "monomers_file" : "",
      "ignore_1b_poly" : [],
      "ignore_2b_poly" : [],
      "ignore_3b_poly" : []
   } ,
   "i-pi" : {
      "port" : 34543,
      "localhost" : "localhost"
   }
}
```

### 9.2.    LAMMPS: lammps.in

```
# DEFINE HEADER
clear
processors * * * map xyz
atom_style full
units metal
boundary p p p
atom_modify sort 0 0.0
neighbor 1.0 bin
```

```
timestep 0.0001

read_data bulk-256.data
mass 1 15.999
mass 2 1.008
variable equil_temp equal 250

# DEFINE CALCULATOR

# DEFINE DUMP/THERMO MODEL
dump initial all custom 200 continue.data id type x y z vx vy vz fx fy fz
thermo_style custom step temp press cpu pxx pyy pzz pxy pxz pyz ke pe etotal vol lx ly lz atoms
thermo_modify flush yes format float %23.16g
thermo 200

# Start movement
velocity all create 100.0 123456

# CONTINUE NPT
fix 2 all nvt temp 200 300 0.05
run 5000
unfix 2

fix 3 all nvt temp 300 300 0.05
timestep 0.0002
run 5000
timestep 0.0003
run 5000
unfix 3

minimize 1.0e-4 1.0e-6 1000 10000

fix 4 all npt temp 300 $(v_equil_temp) 0.05 iso 15 10 0.05
run 5000
unfix 4

fix 5 all npt temp $(v_equil_temp) $(v_equil_temp) 0.05 iso 10 5 0.1
timestep 0.0004
run 5000
unfix 5

fix 6 all npt temp $(v_equil_temp) $(v_equil_temp) 0.05 iso 5 1 0.25
timestep 0.0005
run 5000
unfix 6
```

```
fix 7 all npt temp $(v_equil_temp) $(v_equil_temp) 0.05 iso 1 1 0.5
run 2000000 # 1 ns
write_restart restart.dat
```

### 9.3.  Allegro: allegro.yaml

```
BesselBasis_trainable: true
PolynomialCutoff_p: 12
_jit_bailout_depth: 2
_jit_fuser: fuser1
_jit_fusion_strategy:
- !!python/tuple
  - DYNAMIC
  - 3
allow_tf32: true
append: true
avg_num_neighbors: 56.78007291666667
batch_size: 1
chemical_symbols:
- H
- O
code_commits:
  allegro: f547aa6ced349b0000f328e593d539bb1fa0d3e0
  nequip: 0b02c41cbd30ef9a2f58d95cc3dd41a8beb0ff5d
dataloader_num_workers: 0
dataset: ase
dataset_AtomicData_options:
  r_max: 6.5
dataset_file_name: /mnt/public/tgmaxson/MB-Pol/datasets/1-train.traj
dataset_seed: null
dataset_statistics_stride: 1
default_dtype: float64
device: cuda
e3nn_version: 0.5.1
early_stopping: null
early_stopping_kwargs: null
early_stopping_lower_bounds:
  LR: 1.0e-08
edge_eng_mlp_latent_dimensions:
- 64
edge_eng_mlp_nonlinearity: null
ema_decay: 0.999
ema_use_num_updates: true
embed_initial_edge: true
end_of_batch_callbacks: []
end_of_epoch_callbacks: []
```

```
end_of_train_callbacks: []
env_embed_mlp_latent_dimensions: []
env_embed_mlp_nonlinearity: null
env_embed_multiplicity: 64
equivariance_test: false
exclude_keys:
- magmom
- dipole
final_callbacks: []
global_rescale_scale: dataset_total_energy_std
gpu_oom_offload: false
grad_anomaly_mode: false
init_callbacks: []
initial_model_state: models/train-1-2-5.5-F2/best_model.pth
irreps_edge_sh: 1x0e+1x1o+1x2e
l_max: 2
latent_mlp_latent_dimensions:
- 64
- 64
- 64
latent_mlp_nonlinearity: silu
latent_resnet: true
learning_rate: 0.0005
log_batch_freq: 1
log_epoch_freq: 1
loss_coeffs:
  forces:
  - 1
  - PerSpeciesL1Loss
  stress: 50
  total_energy:
  - 20
  - PerAtomL1Loss
lr_scheduler_factor: 0.9
lr_scheduler_kwargs:
  cooldown: 0
  eps: 1.0e-08
  factor: 0.9
  min_lr: 0
  mode: min
  patience: 400
  threshold: 0.0001
  threshold_mode: rel
  verbose: false
lr_scheduler_name: ReduceLROnPlateau
lr_scheduler_patience: 400
```

```yaml
max_epochs: 50000
max_gradient_norm: .inf
metrics_components:
- - forces
  - mae
  - PerSpecies: true
    report_per_component: false
- - forces
  - rmse
  - PerSpecies: true
    report_per_component: false
- - total_energy
  - mae
  - PerAtom: true
- - total_energy
  - rmse
  - PerAtom: true
- - stress
  - mae
- - stress
  - rmse
metrics_key: validation_loss
model_builders:
- allegro.model.Allegro
- PerSpeciesRescale
- StressForceOutput
- RescaleEnergyEtc
- initialize_from_state
model_debug_mode: false
model_dtype: float32
n_train: 3200
n_train_per_epoch: null
n_val: 500
nequip_version: 0.6.0
nonscalars_include_parity: true
num_basis: 8
num_layers: 2
num_types: 2
optimizer_kwargs:
  amsgrad: false
  betas: !!python/tuple
  - 0.9
  - 0.999
  capturable: false
  differentiable: false
  eps: 1.0e-08
```

```
  foreach: null
  fused: null
  maximize: false
  weight_decay: 0
optimizer_name: Adam
parity: o3_full
per_species_rescale_scales: dataset_per_atom_total_energy_std
per_species_rescale_shifts: dataset_per_atom_total_energy_mean
r_max: 5.5
report_init_validation: true
root: models
run_name: train-1-2-5.5-E
save_checkpoint_freq: -1
save_ema_checkpoint_freq: -1
seed: 12345
shuffle: true
tensorboard: false
torch_version: !!python/object/new:torch.torch_version.TorchVersion
- 2.0.0
```

## 9.4. Ice Structures

### 9.4.1. Ih.POSCAR

```
O H O H O H O H O H O H O H O H O H O H
1.0000000000000000
    7.6035662999999998   0.0000000000000000   0.0000000000000000
   -3.8017831499999981   6.5848815751592502   0.0000000000000000
    0.0000000000000000   0.0000000000000000   7.1429619999999998
  O  H  O  H  O  H  O  H  O  H  O  H  O  H  O  H  O  H  O  H
  1  2  1  2  1  2  1  2  1  2  1  2  1  2  1  2  1  2  1  2
Cartesian
  2.4999094104789328 -0.0000936511531344  0.3983303377235850
  2.5326079986540715  0.0000628638887915  1.3761860659787166
  3.4296104222649024 -0.0013361763358370  0.0973147992619841
 -1.2500351716211660  2.1649389448846534  0.3983303421695457
 -1.2662494802912601  2.1933344433916870  1.3761860948365054
 -1.7159624285904131  2.9694619532059479  0.0973147789535993
  2.5523473463798871  4.4207953030649803  0.3985635503696460
  2.5355734281333855  4.3917420126661435  1.3765541817671054
  2.0877990425729860  3.6161740271772533  0.0964115527508198
  5.1036604866545536  0.0001042860457801  3.9698183425103468
  5.0709462310005602 -0.0000565139431926  4.9476659609766349
  4.1739600847692584  0.0013405112700853  3.6688100480370092
 -2.5517405668503819  4.4199510629401670  3.9698183469747690
 -2.5355221353171031  4.3915398535114640  4.9476659898510054
```

```
-2.0858190707688147   3.6154254328151945   3.6688100277479982
 1.2494362362424243   2.1640870214184997   3.9700167362494621
 1.2662076802299074   2.1931360262772408   4.9480361770304562
 1.7139851003217026   2.9687092677270783   3.6678943416942680
 5.0316510124426506  -0.0008361117262381   6.6515652457377170
 1.7139910550451043   5.8035507516529021   6.9886219922221864
 5.5140050423715916   0.7803592105822275   6.9885984477043310
-2.5165496874448792   4.3571195878318765   6.6515651904655675
-2.0811915030588737   5.1654480244315044   6.9885984457279458
 4.1690267988211289   4.3861351144680585   6.9886219642320544
 1.2858369493964974   2.2271348468531165   6.6508382797009524
 0.3679265776423473   2.2009706469434489   6.9886284523198210
 1.7221331708190284   1.4191190295304230   6.9886284324981691
 2.5719292748026397   0.0008436897788060   3.0800960877498684
 2.0877941254305390   0.7813260357132935   3.4171600233254820
-1.7122324141288137   5.8044993996425447   3.4171959809147792
-1.2852338933799472   2.2277778896633391   3.0800960325100726
 5.8829600257063586   1.4194129561438333   3.4171959789385866
-0.3672488104043462   2.1987458252430008   3.4171599953305472
 2.5159420861270290   4.3577396018307706   3.0793586068349250
 3.4338551461788511   4.3839153540475619   3.4171716586805383
 2.0796545251506120   5.1657635234385797   3.4171716388709785
```

### 9.4.2.  II.POSCAR

```
O H O H O H O H O H O H O H O H O H O H O H
 1.0000000000000000
     7.6051090400000003    0.0000000000000000    0.0000000000000000
    -2.9525527925447750    7.0085744283360345    0.0000000000000000
    -2.9525527925447750   -4.4477024885964269    5.4164617685772445
   O H O H O H O H O H O H O H O H O H O H O H
   1  2  1  2  1  2  1  2  1  2  1  2  1  2  1  2  1  2  1  2  1  2
Cartesian
  1.4443417387770641  -1.2535589438754533   3.6414059260266489
  1.9541832045941954  -1.4687559841235183   4.4426162772205204
  1.9843769951331518  -0.6680879473992505   3.0795892596525003
 -1.1829999891126355   2.3063989025211096   3.3771994991733534
 -1.0067904484800325   1.4780997193603627   3.8680195435053979
 -1.7483077969268939   2.8587493206899928   3.9385104774561372
  2.4804189727257877   2.6668619661974016   0.8139318097881922
  3.2338442375859962   3.1866843329923906   1.1370906178579900
  1.6876660389639457   3.0812494326204511   1.2198021270907646
  0.2559558208651112   3.8144382442176390   1.7750590066880989
 -0.2843893224719931   3.2289516661048285   2.3369020182478506
 -0.2541571430120749   4.0296270017138491   0.9737498523628366
```

```
 2.8830001363907236   0.2544754919819155   2.0392846173114330
 3.4483063258799938  -0.2978792702475582   1.4779774282776634
 2.7067924575961153   1.0827734663314443   1.5484626878497461
-0.7804296955633621  -0.1059791254895529   4.6025590732043407
-1.5338689162110990  -0.6257887335159064   4.2794065134426926
 0.0123115356089499  -0.5203664419241659   4.1966621700099340
 2.0572999075102416   6.0021313392882059   0.9853766352378345
 2.6029989517179066   5.6784185898327273   1.7145372409781274
 2.5025270538284428   5.6327226863936133   0.2025633398378194
 4.8874911563757957  -1.9790574117078199   3.6253088067588126
 4.6599326044912077  -2.2646301224373699   4.5204139393580576
 4.1820803762749525  -2.2644765437405514   3.0206494218805542
-3.6491591331899218   0.8824098750800987   4.6489896820017176
-3.4845372397155643   1.8371551560361334   4.5333603760621140
-3.8188045345191419   0.5381051086043644   3.7615523400395814
-0.3572924063234376  -3.4412636791161964   4.4311008729697221
-0.9029326650744632  -3.1175032683910877   3.7019189741168503
-0.8025236886062305  -3.0718511104664641   5.2139096150970090
-3.1875117011214265   4.5400405493741438   1.7911566547027922
-2.9599079257275074   4.8253494847871918   0.8959714725550736
-2.4820252836840906   4.8253417920958235   2.3957823982355704
 5.3491443205971994   1.6784343673402276   0.7674642082597329
 5.1845437355663551   0.7236835094223912   0.8831190281087397
 5.5187908737923896   2.0227688720444315   1.6548888858944097
```

### 9.4.3.  III.POSCAR

```
O H O H O H O H O H O H O H O H O H O H O H O H
1.0000000000000000
     6.6952420000000004   0.0000000000000000   0.0000000000000000
     0.0000000000000000   6.6952420000000004   0.0000000000000000
     0.0000000000000000   0.0000000000000000   6.7476589999999996
O H O H O H O H O H O H O H O H O H O H O H O H
 1 2 1 2 1 2 1 2 1 2 1 2 1 2 1 2 1 2 1 2 1 2 1 2
Cartesian
 6.0036911801163138   0.6915427591356916   5.0607806269091489
 5.3783723388044278   0.9021185858388525   4.3447172401181700
 5.7931875754316531   1.3168627651094236   5.7767707318415447
 0.6915574663635747   6.0036579744438372   1.6869262366743332
 1.3168399968588038   5.7931269642742667   0.9708629994000891
 0.9020810447706518   5.3783458196178806   2.4029472060996775
 4.0391762872169048   4.0391768220896829   3.3738347085292122
 4.2496356399407391   4.6644725303995758   2.6578036697700758
 4.6645123461174496   4.2496068222255987   4.0898464501282348
 2.6560607211044336   2.6560792559076489   0.0000428899562168
```

2.4455747874367217   2.0307816719971177   6.0316221756131476
2.0307720245610494   2.4456444078084720   0.7160256938183949
4.0901139948875125   1.3131078787967432   3.1359661806446861
3.2647536292615857   1.1103554690549426   3.6157548474865244
4.1246043787731255   2.2862436125950425   3.0843037782081835
2.6051249291943934   5.3821429263439589   6.5097892462669060
3.4304662663208703   5.5849034922777996   0.2419312435233889
2.5706048926109837   4.4089972993094211   6.4581376851508239
4.6607183356976041   5.9527588386388741   1.4490476450986944
4.4579671174297184   0.0828663002044581   1.9288402163673601
5.6338606502267368   5.9182462844027048   1.3973822123224795
2.0345235425892132   0.7424881246196882   4.8228718080584638
2.2372699919928807   6.6123977358091768   5.3026633942542523
1.0613753860959698   0.7769402455320406   4.7712305271791147
5.3821362900234089   2.6051085468889261   0.2378393702260722
5.5849291450192524   3.4304553995210805   6.5057230722312260
4.4089941256641678   2.5706852535772136   0.2895510841184275
1.3131004022090762   4.0900971073784307   3.6116620828193531
1.1103196881379009   3.2647573425961443   3.1319004332720262
2.2862454120551616   4.1246261464588949   3.6633670906344968
5.9527628696094270   4.6607196521811307   5.2985863734400684
0.0828533920452994   4.4579465544827217   4.8188134057301237
5.9183219308489621   5.6338744845759861   5.3502931813807431
0.7424915743763700   2.0345256508777139   1.9247510497702067
6.6123995271283800   2.2373109811441512   1.4449924492616644
0.7769571090808772   1.0613842835421454   1.9764534877874196

### 9.4.4.  Ic.POSCAR

O H O H O H O H O H O H O H
1.0000000000000000
    6.3579999999999997   0.0000000000000000   0.0000000000000000
    0.0000000000000000   6.3579999999999997   0.0000000000000000
    0.0000000000000000   0.0000000000000000   6.3579999999999997
O H O H O H O H O H O H O H
 1  2  1  2  1  2  1  2  1  2  1  2  1  2
Cartesian
 2.3860505313233635   2.3842372304394068   2.3842759240067477
 1.7984213416232857   1.8341686822348866   2.9343274142496019
 1.7984276697322499   2.9343427578561010   1.8341689359204423
 2.3860528380779762   5.5632691696812753   5.5632463912361372
 1.7983930027673425   5.0131676876310083   6.1133411528150541
 1.7984261163214426   6.1133140916607331   5.0131806115375959
 5.5650507181547413   5.5632672108503254   2.3842283614221675
 4.9774108697722141   5.0131616283509155   2.9343395368235221

| | | |
|---|---|---|
| 4.9774336749238959 | 6.1133258235263233 | 1.8341780010818090 |
| 5.5650511310289543 | 2.3842521995195178 | 5.5632468217419406 |
| 4.9773941519903531 | 1.8341660117888692 | 6.1133324198723784 |
| 4.9774003084082166 | 2.9343253552006345 | 5.0131692397364747 |
| 3.9755596792170720 | 3.9737193920429199 | 3.9737439320596608 |
| 3.3878876370630948 | 4.5238273428741405 | 4.5238365336139514 |
| 3.3878900796998948 | 3.4236590115061332 | 3.4236672893887441 |
| 3.9755396377077621 | 0.7947559560558985 | 0.7947314429276184 |
| 3.3878769471743340 | 1.3448437902104688 | 1.3448376985031594 |
| 3.3879465700881917 | 0.2446798844592337 | 0.2446734609619643 |
| 0.7965465678703983 | 0.7947416818316456 | 3.9737302233702785 |
| 0.2089075208656101 | 1.3448459677544937 | 4.5238263839549564 |
| 0.2089427025192342 | 0.2446848616545318 | 3.4236697638804028 |
| 0.7965574460610314 | 3.9737315843939198 | 0.7947492988659901 |
| 0.2088972725194575 | 4.5238311665903250 | 1.3448261104577592 |
| 0.2089594574157444 | 3.4236815120445270 | 0.2446730504064159 |

## 10.    Technical Parameters / Hardware

Models are trained and evaluated on A100 and RTX 4090 graphics processing units (GPUs) with 40 GB and 24 GB of memory, respectively. Models of up to 1024 water molecules are possible on either GPU within these memory constraints, with larger models accessible on the A100. PyTorch version 2.1.2 is used with a system CUDA of version 12.0 on an RTX 4090 workstation and NERSC's Perlmutter cluster. Training time is difficult to evaluate directly due to an exploratory workflow, but the full training time of the model can be estimated at under 2 weeks.

## 11.    References

(1) Marin, F.; Rohatgi, A.; Charlot, S. WebPlotDigitizer, a polyvalent and free software to extract spectra from old astronomical publications: application to ultraviolet spectropolarimetry. arXiv preprint arXiv:1708.02025 2017.

(2) Zhai, Y.; Caruso, A.; Bore, S. L.; Luo, Z.; Paesani, F. A "short blanket" dilemma for a state-of-the-art neural network potential for water: Reproducing experimental properties or the physics of the underlying many-body interactions? The Journal of Chemical Physics 2023, 158 (8), 084111.

(3) Muniz, M. C.; Car, R.; Panagiotopoulos, A. Z. Neural Network Water Model Based on the MB-Pol Many-Body Potential. The Journal of Physical Chemistry B 2023, 127 (42), 9165-9171.

(4) Peterson, A. A.; Christensen, R.; Khorshidi, A. Addressing uncertainty in atomistic machine learning. Physical Chemistry Chemical Physics 2017, 19 (18), 10978-10985, 10.1039/C7CP00375G.

(5) Jeong, W.; Yoo, D.; Lee, K.; Jung, J.; Han, S. Efficient atomic-resolution uncertainty estimation for neural network potentials using a replica ensemble. The Journal of Physical Chemistry Letters 2020, 11 (15), 6090-6096.

(6) Zhu, A.; Batzner, S.; Musaelian, A.; Kozinsky, B. Fast uncertainty estimates in deep learning interatomic potentials. The Journal of Chemical Physics 2023, 158 (16), 164111.

(7) Bukowski, B. C.; Bates, J. S.; Gounder, R.; Greeley, J. First principles, microkinetic, and experimental analysis of Lewis acid site speciation during ethanol dehydration on Sn-Beta zeolites. Journal of Catalysis 2018, 365, 261-276.

(8) Zhong, K.; Yu, C.-C.; Dodia, M.; Bonn, M.; Nagata, Y.; Ohto, T. Vibrational mode frequency correction of liquid water in density functional theory molecular dynamics simulations with van der Waals correction. Physical Chemistry Chemical Physics 2020, 22 (22), 12785-12793.

(9) Amann-Winkel, K.; Bellissent-Funel, M.-C.; Bove, L. E.; Loerting, T.; Nilsson, A.; Paciaroni, A.; Schlesinger, D.; Skinner, L. X-ray and neutron scattering of water. Chemical reviews 2016, 116 (13), 7570-7589.